\documentstyle[12pt]{article}
 \title{Gauge conditions for non-abelian Chern-Simons system consistent
  with equations of motion}

 \author{
  K.Dasgupta\thanks{email: kd@arbornet.org}
M.Sami \thanks{e-mail:sami.ph@jmi.ernet.in}
Pankaj Sharan\thanks{e-mail:pankj.ph@jmi.ernet.in}
  Anupama Mehra
  \\ Department of Physics, Jamia Millia Islamia
  \\ New Delhi 110 025, India}
 \date{}
 
\begin{document}
 \maketitle
 \begin{abstract}
 Complete constraint analysis and choice of gauge conditions consistent with
 equations of motion is done for non-abelian Chern-Simons field interacting
 with N-component complex scalar field.Dirac-Schwinger condition is satisfied
 by the reduced phase-space Hamiltonian density with respect to the Dirac
 bracket.
 \end{abstract}

 PACS number(s): 0.350.Kk, 11.10.Ef \\ \\

 In an earlier work \cite{1} it was shown that for abelian Chern-Simons field
 interacting with complex scalar field,gauge conditions consistent with
 equations of motion can be chosen. It was shown that the choice of gauge
 conditions crucially determines the transformation properties in the
 reduced phase-space and invariance under Lorentz transformations,
 particularly under boosts.Invariance under boosts cannot be demonstrated
  unless the gauge conditions
 are consistent with the equations of motion.

 In the present paper we summarize our results for a similar demonstration for
 the non-abelian Chern-Simons term. This system has been discussed by several
 authors \cite{2,3} but, as far as we know
 proper gauge conditions have not been chosen for this system.
 We show that in this case too  gauge conditions can be chosen which are consistent with equations
 of motion, even though there is no closed form for the reduced phase-space Dirac brackets.
The Dirac-Schwinger conditions are satisfied, at least formally.This ensures
 relativistic invariance. 

  Lagrangian density for non-abelian Chern-Simons field coupled to N-component
  scalar field is of the form
  $$ {\cal L} = [D_\mu\Phi_\alpha]^\dagger [D^\mu\Phi_\alpha]
               +{\kappa\over 4\pi}\epsilon^{\mu\nu\lambda}{\rm Tr}
               [A_\mu\partial_\nu A_\lambda -{2\over 3}
               A_\mu A_\nu A_\lambda]       \eqno(1)
      $$
     where $\Phi $ is N-component scalar field which transforms according
     to fundamental representation of the Lie group SU(N).Here

     $${D_\mu} = {\partial_\mu} - {A_\mu}$$
     $${A_\mu} = ig\, T^a {A_\mu}^a $$

     and $ T^a $ are the fundamental representation matrices.

      $${[T^a , T^b]}_- = i f^{a b c} T^c $$

      $$ {[T^a , T^b]}_+ = d^{a b c} T^c  + {1\over N} {\delta}^{a b} $$

      $$ {\rm Tr}(T^a T^b) = {1\over 2} {\delta}^ {a b} $$

     Where $d^{a b c}$ and $f^{a b c}$ have ususal meaning

      Canonically conjugate momenta for $\Phi$ and $\Phi^\dagger$ are called
     $P$ and $P^\dagger$ where

      $$ P_\alpha = \dot {\Phi_\alpha}^\dagger + i g\, {\Phi_\beta}^\dagger
                         A_0^a T^a_{\beta \alpha}      \eqno(2)     $$

     There are three sets primary constraints

     $$ {\Pi_\rho}^a =  -{\kappa \over 8 \pi} g^2\, \epsilon^{\mu 0 \rho}
                                 {A_\mu}^a              \eqno(3)         $$
  i.e
    $$ \phi_0^a = {\Pi_0}^a \approx 0                  \eqno(4)           $$
    $$ \phi_1^a = {\Pi_1}^a + {\kappa \over 8 \pi} g^2\,A_2^a \eqno(5)           $$
    $$ \phi_2^a = {\Pi_2}^a - {\kappa \over 8 \pi} g^2\,A_1^a \eqno(6)    $$

    of which the latter two are second class.

      Canonical Hamiltonian density is 
      
   \begin{eqnarray*}
    {\cal H}_c & = & {P_\alpha}^\dagger {P_\alpha} +
   (\partial_i\Phi_\alpha^\dagger)(\partial^i\Phi_\alpha) + A_0^a\jmath_0^a
   +A_i^a\jmath_i^a \\
     & + &  g^2 \Phi_\beta^\dagger A_i^a T^a_{\beta \alpha}
       T^b_{\alpha\gamma}+   {\kappa \over 8 \pi} g^2\, \epsilon^ {i j}
       [A_0^a \partial_i A_j^a + A_i^a\partial_j A_0^a] \\
     & + & i {\kappa \over 8 \pi} g^3\, \epsilon^ {i j} A_0^a A_i^c A_j^b
      [if^{a b c} + d^{a b c}]
      \end{eqnarray*}

    where

    $$ \jmath_0^a = ig\, [P_\alpha \Phi_\beta - P_\alpha^\dagger \Phi_\beta^\dagger]
                    T^a_{\beta \alpha}     \eqno(7)                      $$

    $$ \jmath_i^a = -ig\, [(\partial_i\Phi_\alpha^\dagger)\Phi_\beta -
                     \Phi_\beta^\dagger (\partial_i \Phi_\alpha)] T^a_
                     {\beta \alpha}          \eqno(8)                 $$

    Adding the primary constraints to the canonical Hamiltonian 

    $$ H_p = {\int d^2 x{\cal H}_c}  + {\int d^2 x(v^0\phi_0 + v^1\phi_1
               +v^2\phi_2)} $$

     and using the basic Poisson's brackets

      $$\dot\phi_0^a = \{\Pi_0^a (\vec x ,t) , H_p \} \approx 0 $$

     gives a secondary constraint $\phi_3^a$,

      $$ \phi_3^a ={\jmath_0}^a + {\kappa \over 4 \pi} g^2\,
                                   \epsilon^{i j}
                                   [\partial_i A_j^b -{g \over 2} A_i^c A_j^e
                                            f^{b c e}] \approx 0
                              \eqno(9)               $$

    this can be made into a first class constraint by choosing

        $$\psi^a =  \partial_1 \phi_1^a + \partial_2 \phi_2^a + \phi_3^a \approx 0
                                      \eqno(10)              $$

   Conservation in time of two second class constraints $\phi_1^a$ and $\phi_2^a$
    determines  $v^1$ and $v^2$ which can be
   substituted back.

  To choose gauge conditions for the two first class constraints $\phi_1^a$
 $\psi$  we try $\nabla \cdot A^a = 0$. This requires,
  for consistency with equations of motion that

\begin{eqnarray*}
{d\over dt} (\nabla \cdot A^a) & = & -  {4\pi \over \kappa g^2}
        ({ \partial_1 \jmath_2}^a -  { \partial_2 \jmath_1}^a )
        - \nabla^2 A_0^a \\  & - &  {4\pi \over \kappa g^2}
    [\partial_1 (\Phi_\beta^\dagger T^a_{\beta \alpha} T^b_{\alpha \gamma}A_2^b \Phi_\gamma)
 -   \partial_2 (\Phi_\beta^\dagger T^a_{\beta \alpha} T^b_{\alpha \gamma}A_1^b \Phi_\gamma)] \\ 
& + &  {g \over 2} [ \partial_1 (A_0^e A_1^b) -  \partial_2 (A_0^e A_2^b)] f^{e b a }
  \end{eqnarray*}
   should be equated to zero.
   Thus

\begin{eqnarray*}
  \nabla^2 A_0^a  & = &  - {4\pi \over \kappa g^2}
        ({ \partial_1 \jmath_2}^a -  { \partial_2 \jmath_1}^a )-
           {4\pi \over \kappa g^2}
    [\partial_1 (\Phi_\beta^\dagger T^a_{\beta \alpha} T^b_{\alpha \gamma}A_2^b \Phi_\gamma)
 -   \partial_2 (\Phi_\beta^\dagger T^a_{\beta \alpha} T^b_{\alpha \gamma}A_1^b \Phi_\gamma)] \\ 
& + &  {g \over 2} [ \partial_1 (A_0^e A_1^b) -  \partial_2 (A_0^e A_2^b)] f^{e b a }
  \end{eqnarray*}

  There is no closed form solution, but it can be solved iteratively as a
  formal series \cite{4} . The result is
$$\chi_0^a =  \ A_0^a +  {\lambda \over 2\pi} {\int d^2 y} ln |x-y|
              (\partial_1 \tilde \jmath_2^a - (\partial_2 \tilde \jmath_1^a )
              + \cdots       \eqno(11)                    $$

 Where

 $$ \lambda = {4\pi \over \kappa g^2} $$

 $$ \tilde \jmath_1^a = \jmath_1^a + {\Phi_\beta^\dagger T^a_{\beta \alpha} T^b_{\alpha \gamma}A_1^b \Phi_\gamma} $$

 $$ \tilde \jmath_2^a = \jmath_2^a + {\Phi_\beta^\dagger T^a_{\beta \alpha} T^b_{\alpha \gamma}A_2^b \Phi_\gamma} $$

Thus $\chi_0^a \approx 0 $ is the first gauge condition and
 $ \chi^a = (\nabla \cdot A^a)  \approx 0 $ is the second gauge condition.

  Using $ \phi_1^a $ and $ \phi_2^a $ we can form Dirac brackets. Next we can do
  the same with $\psi^a$ and $\chi^a$ and put them strongly equal to zero

 Therefore
$$ \nabla \times A^a -g \epsilon^{i j} A^c_i A^e_j f^{a c e} +\lambda \jmath^a_0 = 0
                              \eqno(12)                     $$
$$ \nabla \cdot A^a = 0                    \eqno(13)            $$

  These can solved as

 $$ A^a_1=-\partial_2 B^a \quad {\rm and} \quad A^a_2= \partial_1 B^a $$

Therefore

$$
\nabla^2 B^a + 2g(\partial_2 B^e )(\partial_1 B^c) f^{a e c} +
    \lambda \jmath^a_0 \approx 0             \eqno(14)                     $$
which can be solved for $B^a$ formally just as $A_0^a$ was solved.

Reduced phase-space Hamiltonian density is of the form
\begin{eqnarray*}
\Theta^{0 0} &=& P_\alpha^\dagger P_\alpha + (\partial_i \Phi_\alpha^\dagger)
(\partial_i \Phi_\alpha) + A_i^a \jmath_i^a\\ &+&
g^2\Phi_\beta^\dagger A_i^a  T^a_{\beta \alpha} T^b_{\alpha \gamma} A_i^b \Phi_\gamma
\end{eqnarray*}

Dirac-Schwinger equal time  condition in reduced phase-space is

$$\{ \Theta^{0 0}(x), \Theta^{0 0}(y) \} = - \partial_k \delta(\vec x - \vec y)
[\Theta^{0 k}(x) + \Theta^{0 k}(y)] $$
$$ \Theta^{0 k} = P_\alpha \partial^k \Phi_\alpha + P_\alpha^\dagger
\partial^k \Phi_\alpha^\dagger + A^{k c} \jmath_0^c     \eqno(15)          $$

  The transformation for fields can be done through the generators

  $$ P^k = {\int d^2 x}\Theta^{0 k}         \eqno(16)                       $$
  $$ M^{\mu \nu} = {\int d^2 x} [x^\mu \Theta^{0 \nu} - x^\nu \Theta^{0 \mu}]
                    \eqno(17)                          $$

   We therefore see that gauge conditions
  compatible with equations of motion can be chosen for non-abelian case also
  giving qualitatively similar results to the abelian case.

{\em Work by one of us (KD) was supported by the Indian CSIR grant 
no.9/466(30)/96-EMR-I.}

 \end{document}